\documentclass[fleqn,usenatbib]{mnras}

\usepackage{newtxtext,newtxmath}

\usepackage[T1]{fontenc}
\usepackage{ae,aecompl}

\usepackage{graphicx}	
\usepackage{amsmath}	
\usepackage{natbib}
\usepackage[parfill]{parskip}
\usepackage{enumitem}
\usepackage[mathscr]{euscript}

\usepackage{multirow}
\usepackage[FIGTOPCAP]{subfigure}
\usepackage{media9}

\numberwithin{equation}{section}

\title[Ionized Bondi Accretion]{Stability of an Ionization Front in Bondi Accretion}

\author[Keto \& Kuiper]{
Eric Keto$^{1}$\thanks{E-mail: eketo@cfa.harvard.edu (EK)}
Rolf Kuiper$^{2}$\thanks{E-mail: kuiper@uni-heidelberg.de(RK)}
\\
$^{1}$Institute for Theory and Computation, Harvard College Observatory, 60 Garden St., Cambridge, MA 02138\\
$^{2}$Zentrum f{\"u}r Astronomie der Universit{\"a}t Heidelberg, Institut f{\"u}r Theoretische Astrophysik, Albert-Ueberle-Stra{\ss}e 2, 69120 Heidelberg, Germany\\
}

\date{Accepted XXX. Received YYY; in original form ZZZZ}

\pubyear{2020}

\begin{document}
\label{firstpage}
\pagerange{\pageref{firstpage}--\pageref{lastpage}}
\maketitle

\begin{abstract}
Spherical Bondi accretion is used in astrophysics as an approximation to investigate many types of accretion processes.
Two-phase accretion flows that transition from neutral to ionized have observational support in high-mass star formation, and have application
to accretion flows around any ionizing source, but the
hydrodynamic stability of two-phase Bondi accretion is not understood. With both semi-analytic and fully numerical methods we find that
these flows may be stable, conditionally stable or unstable depending on the initial conditions. The transition from an R-type to
a D-type ionization front plays a key role in conditionally stable and unstable flows.
\vskip 0.25truein
\end{abstract}



\section{Introduction}

Shortly after Bondi's explanation  \citep{Bondi1952} of his eponymous accretion flow, \citet{Mestel1954}  
described a Bondi accretion flow through a static ionization front as might exist around a compact ionizing 
source. The Rankine-Hugoniot equations for the discontinuity across the ionization front
allow solutions for supersonic or subsonic relative velocities of the front with respect to the neutral accretion flow. However, there is a
range of relative velocities, approximately around the sound speed of the ionized gas, for which the jump conditions
result in an unphysical square root of a negative number. Thus at that time, the applicability of Mestel's model was not clear. 
A year later \citet{Savedoff1955} resolved 
the problematic jump conditions with a double discontinuity made up of a shock front preceding the
ionization front.  

Observations of radio recombination lines from the ultracompact HII region, G10.6-04, matching the morphology and inward velocity
of the surrounding molecular accretion flow \citep{Keto2002a,Keto2002b} suggest the relevance of Mestel's model with a molecular accretion flow
passing through the ionization front at the HII region boundary and continuing to the star as an ionized accretion flow. 
This interpretation recasts the role of an HII region around a massive accreting protostar from a disruptor to a
participant in the accretion flow. The model is also applicable
to rotationally-flattened accretion flows where the accretion is confined about the mid-plane of the rotating system \citep{Keto2007}. 

The temporal evolution of a two-phase accretion flow is described in \citet{Keto2002b} as a continuous transition through steady-state solutions parameterized by
the ionizing luminosity or equivalently the position of the ionization front. 
This description is adequate assuming that the 
time scale for the ionized flow 
to adjust to a 
new steady-state solution, the fluid-crossing time, is short compared to 
the time scale
for changes in the velocity of the ionization front.
However, the velocity of the ionization front depends on
 the ionizing luminosity and the ionized and neutral gas densities (\S \ref{ionization}). These are both  independent
 of the gas velocity in the sense that Bondi's solution contains an arbitrary scaling factor for the density. 
Therefore the assumption of a steady-state flow in the ionized gas  is not guaranteed.
Recent numerical hydrodynamic studies found no stable solutions 
and suggested that the dynamics in this model may be inherently unstable 
 \citep{Vandenbroucke2019}.  
 
Owing to its simplicity and flexibility rather than astrophysical realism 
\footnote{For example, both the self-gravity and angular momentum of the accretion flow must be negligible.}, 
the spherically-symmetric Bondi accretion flow has been a useful
approximation for investigations of a wide range of accretion phenomena. The stability  of
two-phase Bondi accretion flows is therefore worth understanding in order to extend
the applicability of this widely used approximation.

In this paper, we revisit the stability of an ionization front in a Bondi accretion flow through two different
numerical methods. We follow \citet{Keto2020} and use the method of characteristics to solve the partial differential equation for time-dependent
Bondi accretion within the HII region as a system of coupled ordinary differential equations (\S \ref{SODE}). We also model the 
entire accretion flow from neutral to ionized, passing through the ionization front, with a numerical hydrodynamic
simulation gridded in spherical geometry appropriate to the problem (\S \ref{NHS}). 

In addition to the unstable solutions found in \citet{Vandenbroucke2019}, we find stable solutions with damped oscillations and conditionally stable solutions with
oscillations of limited amplification. The behavior depends on the relationship between two time scales: the time scale for the HII region
to change density, essentially the fluid-crossing time; and the time scale for the ionization front to move to the new position of radiative equilibrium
with the new density. The oscillations result as the front overshoots its equilibrium position.

\section{Solution by Ordinary Differential Equations}\label{SODE}

\subsection{Bondi accretion}\label{bondi}

Following 
\citet{Bondi1952} 
and 
\citet{Parker1958}, the Euler equation in spherical symmetry for
an isothermal gas with  $\partial P / \partial \rho = a^2$ for sound speed $a$, and
a gravitational force from a constant mass, $M$, is \citep{Keto2020}
\begin{equation}\label{tdbpeq}
\frac {\partial \tilde{u}} {\partial \tilde{t}} = -\tilde{u} \frac{\partial \tilde{u}}{\partial \tilde{x}} - \frac{a^2}{\tilde{\rho}} \frac{\partial \tilde{\rho}}{\partial \tilde{x}} - \frac{GM}{\tilde{x}^2}
\end{equation}
where the tilde indicates a variable with dimensional units.
This can be written in non-dimensional form with the definitions,
\begin{equation}\label{scaling}
\tilde{x} = \bigg(\frac{GM}{a^2}\bigg) x , \ \ \tilde{u} = au ,\ \  \tilde{t} = \bigg(\frac{GM}{a^3}\bigg) t , \ \ \rm{and } \ \ \tilde{\rho} = \tilde{\rho_0}\rho .
\end{equation}
where $\tilde{\rho_0}$ is an arbitrary density.
With these substitutions, equation \ref{tdbpeq} is,
\begin{equation}\label{ndbpeq}
\frac {\partial u}{\partial t} = -u \frac {\partial u} {\partial x} - \frac {1} {\rho}   \frac {\partial \rho}{\partial x} - \frac{1}{x^2}.
\end{equation}
The density may be eliminated with the help of the non-dimensional continuity equation,
\begin{equation}\label{continuity}
\rho = \lambda x^{-2}u^{-1}
\end{equation}
where $\lambda$ is the accretion rate. Then,
\begin{equation}\label{tdndeq}
\frac{\partial u}{\partial t} = \bigg( \frac{1}{u} - u\bigg) \frac {\partial u}{\partial x} + \bigg(\frac{2}{x} - \frac{1}{x^2}\bigg).
\end{equation}
In steady state, the time derivative on the left-hand side is zero, and
the variables can be separated
and integrated,
\begin{equation}\label{ss}
 \mathscr{L} = \log |u| - \frac{1}{2} u^2  + 2\log |x| + \frac{1}{x} .
\end{equation}
In non-dimensional units, the constant of integration $\mathscr{L}=\log \lambda$ is equivalent to the energy and
with a simple non-linear scaling to the 
the mass accretion rate.

There is only one solution that transitions from subsonic to supersonic, the Bondi accretion flow  
with $\mathscr{L} = \mathscr{L}_C = 3/2 - 2\log 2$.
This solution passes through a
critical point, $(x_c,u) = (\frac{1}{2},1)$ or $(\tilde{x},\tilde{u}) = (GM/(2a^2), a)$. 
The coefficients (eqns. \ref{scaling}) used
for the non-dimensional scaling are then, respectively in the order listed, twice the radial distance of the transonic critical point,
the sound speed, and twice the sound crossing time to the critical point. The density has been 
eliminated from the solution, and its scaling is arbitrary.

\subsection{Ionization}\label{ionization}

The neutral accretion flow with velocity, $u_1(x)$, follows the unique transonic steady-state 
solution until it
reaches the ionization front at position $x_f(t)$. Here the flow is ionized, slowed and compressed and then 
follows a solution of the partial differential equation, \ref{tdndeq}, for velocity
$u(x,t)$, and
energy constant, 
\begin{equation}\label{energyconstant}
\mathscr{L}= \left( \frac{1}{x_f} + 2\log{x_f} \right) - \left(\frac{1}{2}u_2^2(x_f) - \log{ | u_{2}(x_f) | } \right)
\end{equation}
where $u_{2}(x_f)$ is the velocity of the ionized gas at the position of the front, $x_f$.

\begin{figure}
\includegraphics[width=3.25in]{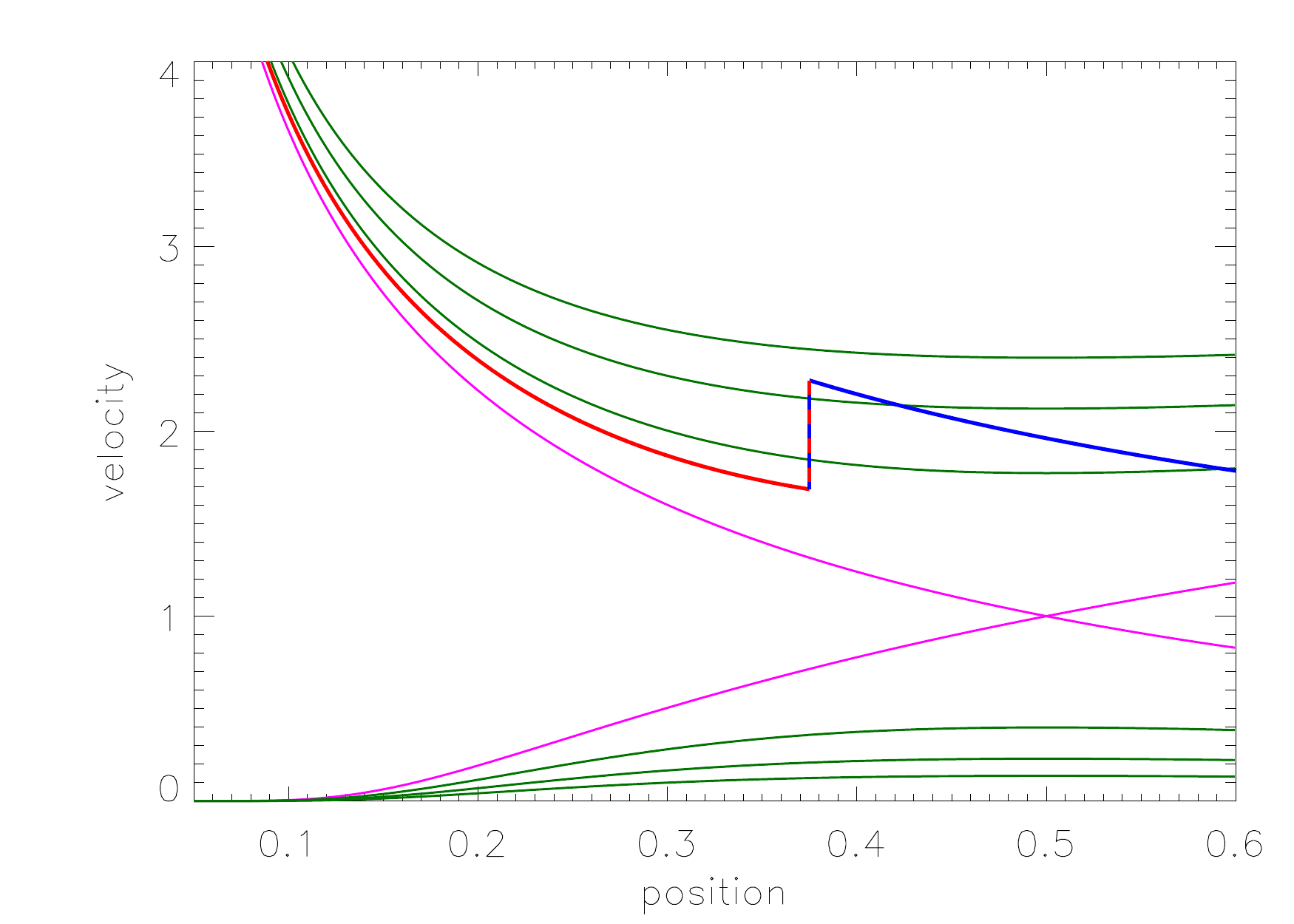}
\caption{
Velocity, $u$, as a function of position, $x$, in non-dimensional units scaled on the sound speed of the ionized
gas. The pink lines are the transonic Bondi accretion
flow and the Parker wind, crossing at the critical point $(x_c,u) = (1/2,1)$. The green
lines show some examples of subsonic and supersonic steady-state solutions.
The  blue and red lines show the initial conditions. The neutral flow (blue line) passes through
the ionization front to continue inward as an ionized accretion flow (red line). The ionization front
is at position 3/8. }
\label{initial_plot}
\end{figure}

In the following, the subscripts $1$ and $2$ indicate the neutral and ionized sides of the front, respectively, and
$\hat{u}$ a
relative velocity of the gas with respect to the front,  $\hat{u} = \dot{x}_f - u$, assuming the gas velocity
is negative for accretion. Then the relative velocity,  $\hat{u}_{2}$, is found from the
Rankine-Hugoniot or jump conditions for continuity and
momentum  in non-dimensional form,
\begin{equation}\label{jump1}
{\rho}_{2}  \hat{u}_{2} = {\rho}_{1} \hat{u}_{1}
\end{equation}
\begin{equation}\label{jump2}
{\rho}_{2}  (1 + \hat{u}_{2}^2) = {\rho}_{1} (\beta^2 + \hat{u}_{1}^2 )
\end{equation}
where $\beta = a_1/a_2$, 
and the non-dimensional scaling coefficients, equations \ref{scaling}, are defined with sound speed, $a_2$.
The jump conditions are combined,
\begin{equation}
1 + \hat{u}_{1}^2  \bigg( \frac {\rho_{1}}{\rho_{2}} \bigg)^2 - \bigg( \beta^2 + \hat{u}_1^2 \bigg) \bigg ( \frac{\rho_{1}} {\rho_{2}} \bigg)= 0 
\end{equation}
and solved as a quadratic in the ratio, $\rho_{1}/\rho_{2}$. The velocity $\hat{u}_{2}$ in terms of the known velocity $\hat{u}_{1}$ is,
\begin{equation}\label{eta1}
\hat{u}_{2} = \eta \pm ( \eta^2 - 1 )^{1/2} 
\end{equation}
with
\begin{equation}\label{eta2}
\eta = \frac {1}{2\hat{u}_{1}} \bigg( \beta^2 + \hat{u}_{1}^2 \bigg) .
\end{equation}
The positive branch, conventionally called R-type, applies when the relative velocity, $\hat{u}_1$, is supersonic,
$\hat{u}_1 \geq u_R = 1 + \sqrt{1-\beta^2} \approx 2$, and the negative branch, D-type, when subsonic,
$\hat{u}_1 \leq u_D = 1 - \sqrt{1-\beta^2} \approx \beta^2/2$. Relative velocities between $u_D$ and $u_R$ result
in an unphysical negative argument of the square root in equation \ref{eta1}. 
The classification of ionization fronts is
explained more fully in textbooks such as \citet{Shu1991} or \citet{Spitzer1978}.
The initial conditions are shown in figure \ref{initial_plot}.

\subsection{Velocity of the Ionization Front}\label{vfront}

The velocity of the ionization front is set by the requirement of equality between the number of neutral atoms flowing into the front
and the number of ionizing photons arriving at the front, both per unit time. The former depends on the 
relative velocity of the front and the neutral accretion flow, as well as the neutral density. The latter
depends on the stellar rate of emission, $\tilde{ J}_*$, less the rate of recombinations in the HII region.
\begin{equation}\label{frontv0}
\left( \frac{d \tilde{x}_f} {dt} -\tilde{ u}_{1}(x_f) \right) r\pi \tilde{x}_f^2 \frac{ \tilde{\rho}_{1}(x_f)} {m_H}
= \tilde{J}_* - \int_{\tilde{x}_*}^{ \tilde{x}_f} \alpha \frac{\tilde{\rho}^2(x^\prime)} {m_H^2} 4\pi x^{\prime 2} dx^\prime ,
\end{equation}
with $m_H$ the mass of the hydrogen atom, and
the type-1 recombination rate coefficient is $\alpha = 3\times 10^{-13}$
cm$^{3}$ s$^{-1}$.  Ionization equilibrium is
explained more fully in textbooks such as \citet{Shu1991} or \citet{Spitzer1978}.

This equation can be written in non-dimensional form as,
\begin{equation}\label{frontv1}
\left( \dot{x}_f - u_{1}(x_f) \right) x_f^2 \rho_{1}(x_f) = J_*  - \int _{x_*}^{x_f} \rho^2 x^{\prime 2} dx^\prime
\end{equation}
or using the continuity equation \ref{continuity} to eliminate the density as,
\begin{equation}\label{frontv2}
\left( \dot{x}_f - u_{1}(x_f) \right) = \frac {u_{1}(x_f)} {\lambda} \left( J_*  - \lambda^2\int _{x_*}^{x_f} u^{-2}(x^\prime) x^{\prime -2} dx^\prime \right) .
\end{equation}
In deriving the non-dimensional form, we set the density scaling coefficient in eqn. \ref{scaling},
\begin{equation}
\left( \frac{m_H}{\alpha \rho_0} \right) \left( \frac{a^3} {GM} \right) = 1.
\end{equation}
The two terms in parentheses are the ratio of the recombination time to the crossing time.
Also,
\begin{equation}
{J}_* = \left[ \frac{1}{a} \frac{1}{4\pi} \left( \frac{a^2}{GM} \right) \frac{m_H}{\rho_0} \right] \tilde{J}_*.
\end{equation}
The right side is the ratio of 
the rate of ionizing photons from the star to the rate of neutral atoms 
through the critical point of the ionized Bondi accretion flow.

\subsection{Time-dependent solution}\label{timedependence}

The time-dependent evolution
can be calculated from the partial differential
equation (PDE) for the Bondi accretion flow, \ref{tdndeq}, combined with the equation for the 
velocity of ionization front \ref{frontv2}. 
Following the method of characteristics as explained in \citet{Keto2020}, 
the PDE may be converted into two first order ordinary differential equations (ODEs),
\begin{equation}\label{ode1}
\frac{dx}{dt} = u - \frac{1}{u}
\end{equation}
\begin{equation}\label{ode2}
\frac{du}{dt} = \frac{2}{x} - \frac{1}{x^2}
\end{equation}
to be solved with initial values, $u(t,x) = u_0(x)$ at $t=0$.
With equation \ref{frontv2} for the position of the ionization front and its initial condition 
we  have a system of three coupled
ODEs. The neutral flow upstream of the HII boundary layer, either a simple ionization front or
a combination of an ionization and a shock front, is independent of time, 
and we solve only for the ionized
portion of the accretion flow and the motion of the front. Our numerical
hydrodynamic simulations described later in \S \ref{NHS} model the flow in the ionized zone and in both the pre- and post-shock
neutral zones. 

\begin{figure*}
\begin{tabular} {p{3.0in}c}
\includegraphics[width=3in]{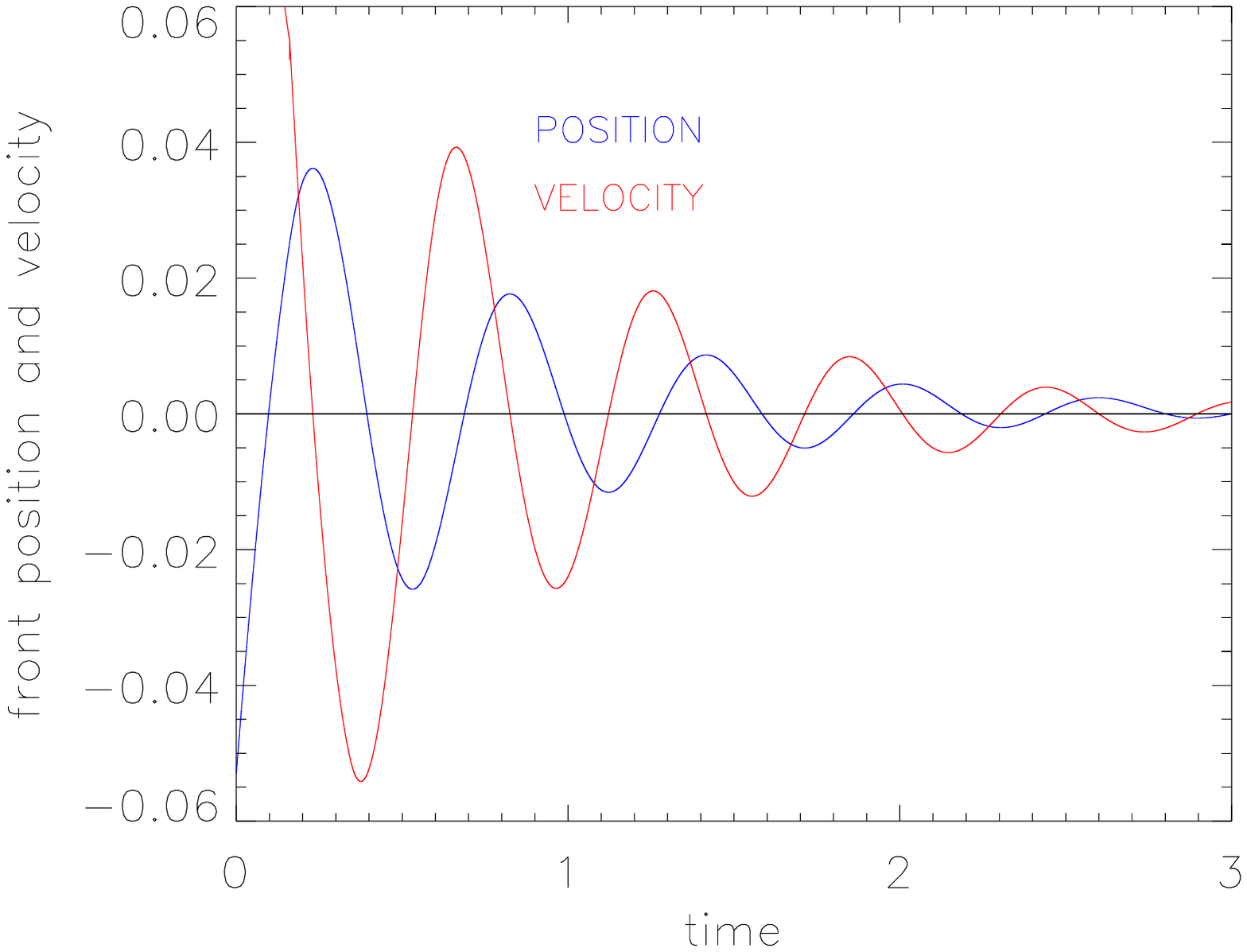}
&
\includegraphics[width=3in]{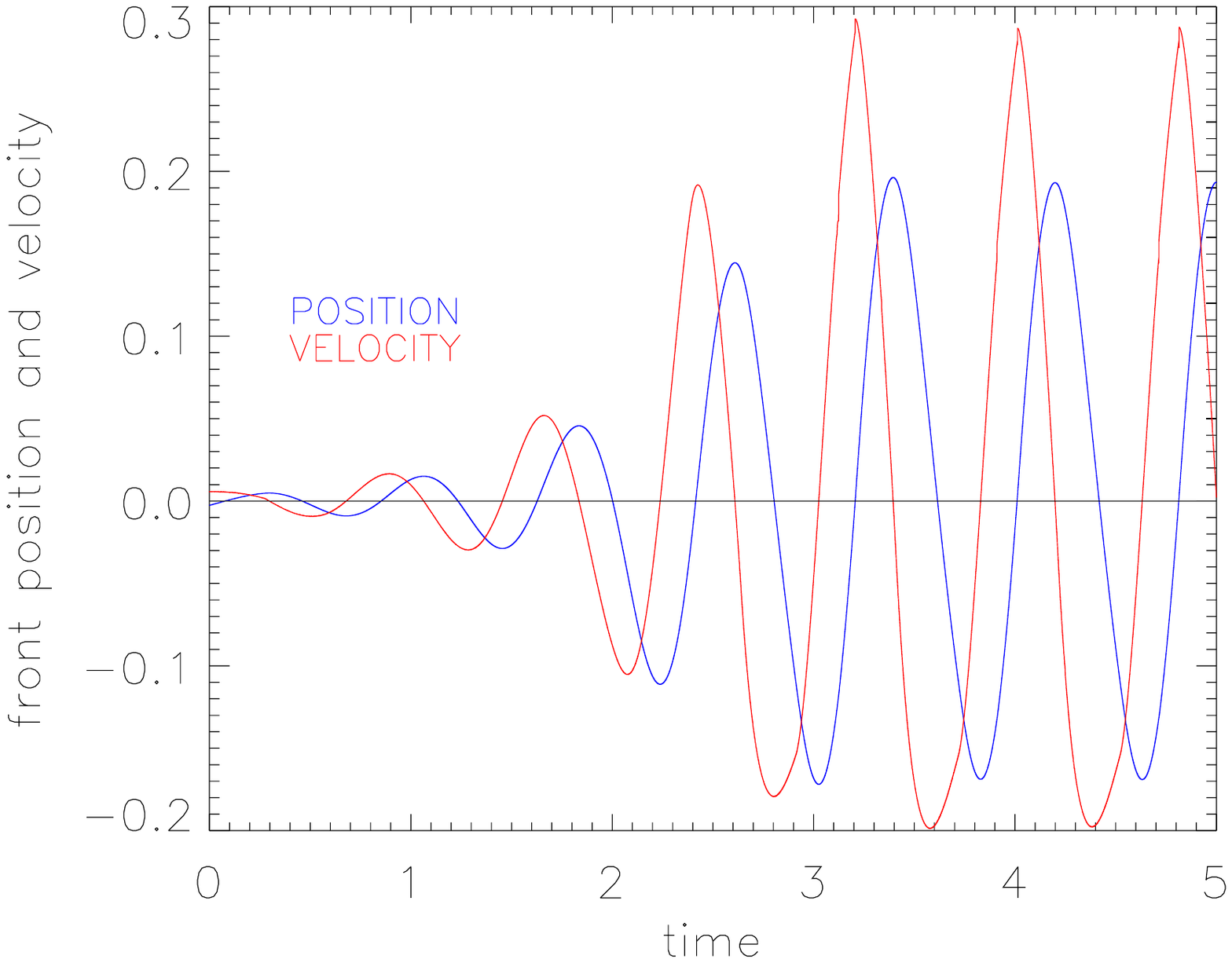}
\end{tabular}
\caption{
Velocity of the ionization front (red) and its position about equilibrium (blue), both shown relative to their
respective equilibria and both in
non-dimensional units.
The initial position of the front is 3/8 (left) and 7/16 (right) and initially out of equilibrium with the stellar flux by 2\% (left)
and 0.1\% (right).
The amplitudes of the oscillations are limited as shown in the figure (right) and continue at this level indefinitely.
}
\label{position1}
\end{figure*}

The system
of coupled ODEs is an initial value problem (IVP) whereas we want to solve a boundary value problem (BVP) 
for the
flow between the surface of the star
and the time-dependent position of the ionization front. Accordingly, at each time step
we remove from the calculation the characteristics represented by equation \ref{ode1} that
are advected past the inner boundary, and we add new characteristics to represent
the gas flowing inward through the outer boundary. The initial velocities for the new
characteristics are determined from equation \ref{eta1} for the combined jump conditions.
Example solutions with damped and amplified oscillations are shown
in figure \ref{position1}. 

\begin{figure*}
\begin{tabular} {p{3.0in}c}
\includegraphics[width=3in]{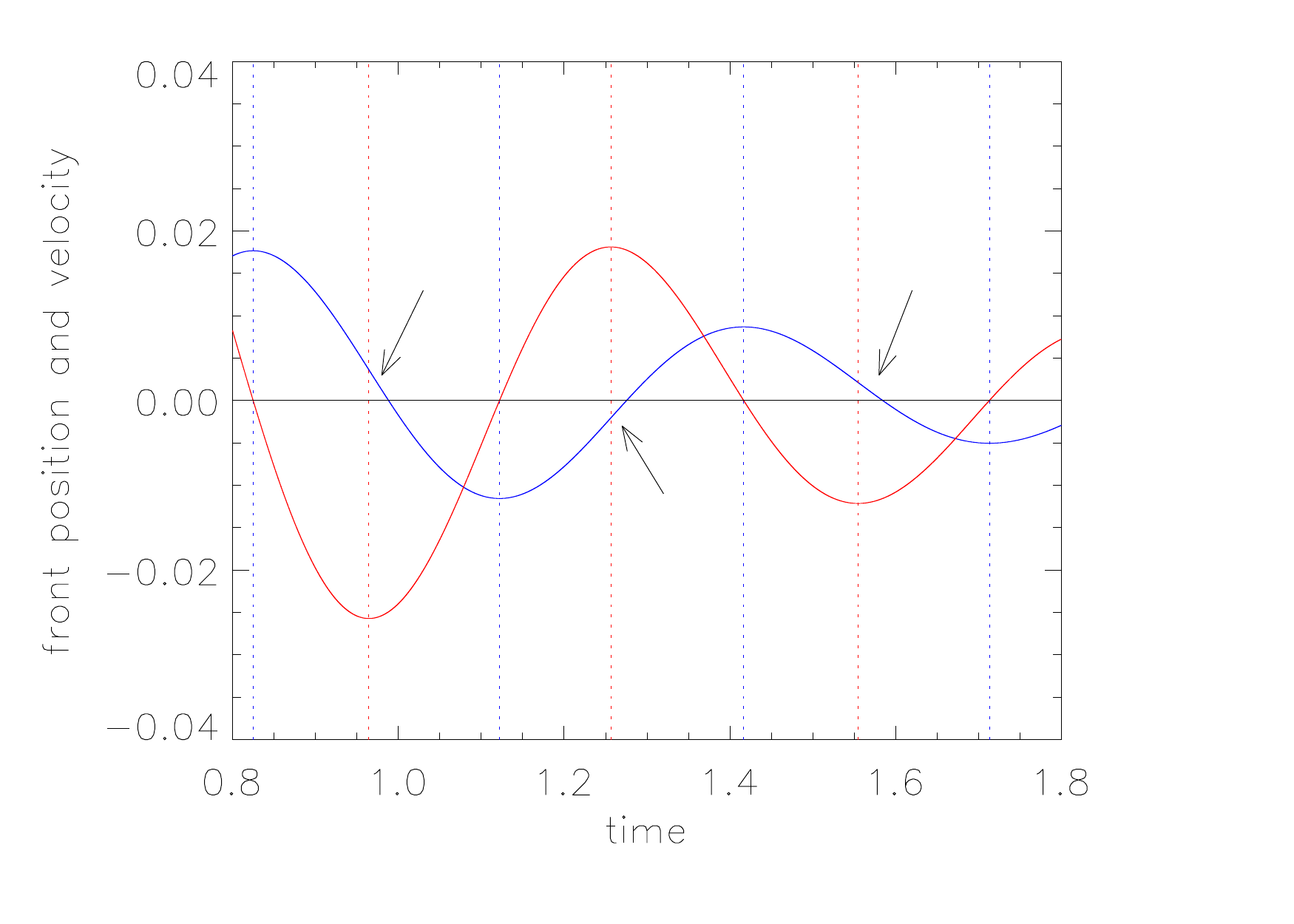}
&
\includegraphics[width=3in]{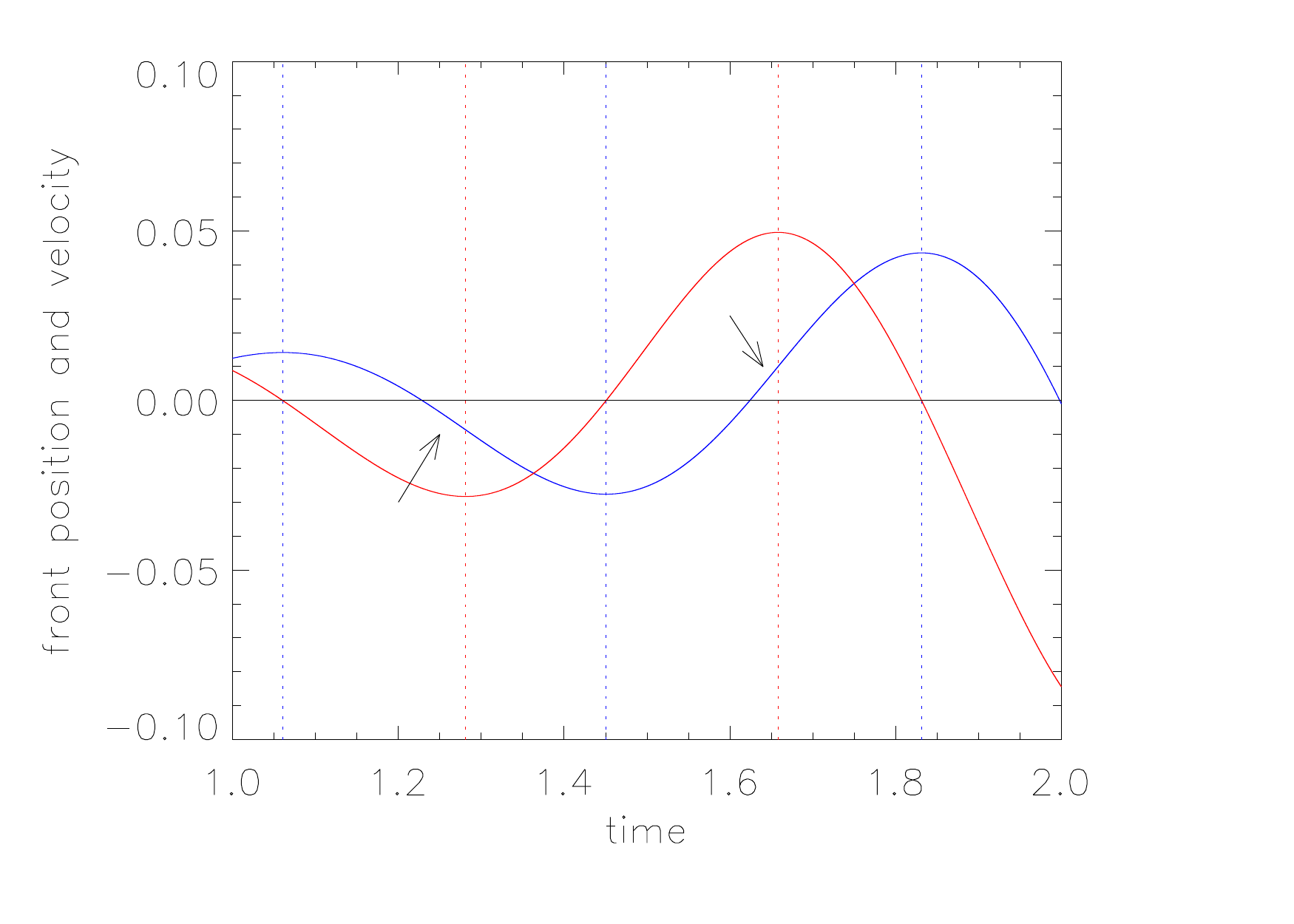}
\end{tabular}
\caption{
Detail of figure \ref{position1}. The arrows point to locations where it is easy to see that the change in the sign of the velocity
does not occur exactly at the time that the position of the front reaches its maximum or minimum from equilibrium.
In the case of damped oscillations (left), the oscillations of velocity lag those in position. If the oscillations in velocity lead those
in position (right), the oscillations are amplified.}
\label{position_detail}
\end{figure*}

\subsection{Explanation} 

The oscillations are damped or amplified depending on the relationship between the
phase
of the oscillations of the velocity of the front and the phase of the oscillations of its position.
The chain of causality involves several steps.
The accretion flow is supersonic and downstream changes in the ionized portion of the flow do not propagate to the ionization
front. The neutral flow upstream is steady by assumption. The movement of the ionization front  
is therefore determined by the ionization
balance  which depends on the ionized gas density through the total recombination rate expressed
by the integral in equation \ref{frontv1}. 

The complexity arises because the motion of the front to establish a new equilibrium
changes
the relative velocity of the front with respect to the steady, neutral flow and thereby
the compression across the front and the density of the gas entering the HII region.  
However, equilibrium requires a static front. 
Therefore, 
the ionized 
density entering the HII region through the moving front is not the density required for equilibrium. 
Advection then propagates this non-equilibrium density through
the HII region, affecting the density-dependent recombination rate, 
the ionizing flux at the front, and in turn the velocity of the front. Generally, the front position overshoots its equilibrium 
resulting in 
oscillations around the equilibrium position.  

If the front is initially relatively close to the star, i.e. the HII region is small compared to the Bondi transonic radius
(e.g. $x_f/x_c = 3/8$ and $1/5$ in our example ODE and fully numerical solutions, respectively), 
then the advection time scale is short,  and
the oscillations of the position of the front lag the oscillations in the velocity of the front. In this case, the overshoot
works to dampen the oscillations  which
decay relatively quickly to the equilibrium position.  If the HII region is larger (e.g. $x_f/x_c = 7/16$ and $1/2$
in our example ODE and fully numerical solutions, respectively), the advection time
scale is longer, and the oscillations of the position of the front lead those of the
velocity of the front, and the overshoot is amplified. 
Figure \ref{position_detail} illustrates the two cases. The arrows point to locations on the
figure where it is easy to see that the oscillations of position and velocity are not quite $180^\circ$ out of phase. 
In the case of damped oscillations the change in the sign of the velocity occurs just after the front has reached its
furthest deviation from equilibrium for each cycle, whereas the opposite occurs in the case of amplified oscillations.
In the two cases, the time differences either work to dampen or amplify, respectively, the oscillation in the next cycle.

For any set of initial conditions there is a critical radius for the initial position of the ionization front that determines
whether the evolution is stable or unstable. However, the critical radius is not universal but depends on the
initial conditions, in particular on the locations of the inner and outer boundaries. 
At least for the moment, this critical radius must be found numerically by following the evolution.

\subsubsection{Development of stable oscillations}\label{StableOscillations}
If the oscillations are amplified, an ionization front starting from an initial position where $\hat{u}_1  \geq u_R$ 
may may evolve such that $\hat{u}_1 = u_R$ even if the ionizing flux, $J_*$ is held constant. 
This generally occurs in the phase of the oscillation cycle when the position of the front is accelerating 
inward in the same direction as the
accretion flow and reducing the relative velocity of the front and the flow.
If the relative velocity, $\hat{u}_1$, were to fall below the critical velocity, $u_R$, the
argument of the square root in equation \ref{eta1} would become negative resulting in unphysical jump 
conditions across the ionization front. 
This situation is resolved by the development of a shock front on the upstream side of
the ionization front as suggested by \citet{Savedoff1955}.
An ionization front at the critical relative velocity, $u_R$, is equivalent to a double front consisting of a D-type ionization 
front with a coincident and co-moving isothermal shock.   In this equivalence, the shock slows the neutral gas entering the ionization
front to the relative velocity, $u_D$, according to the jump conditions across an isothermal shock, 
\begin{equation} 
\hat{u}_{1s} = \beta^2/(u_1 - \dot{x}_s) ,
\end{equation}
where $\hat{u}_{1s}$ is the relative velocity of the neutral gas behind the shock front and $\dot{x}_s$ is the velocity of the
shock front. 

Our ODE model consists of two zones, the undisturbed neutral accretion flow and the HII region,
and does not include a post-shock, compressed, neutral layer. If we assume that the post-shock layer is thin so that
$\hat{u}_{1f} \approx \hat{u}_{1s}$, then the ionization and shock
fronts need not be coincident. 
If we further assume that the compression in the post-shock layer results in the minimum pressure required 
to move the shock front ahead of the ionization front at the minimum velocity required for allowable jump 
conditions,
then the relative velocity of the ionized gas behind the ionization front remains at the critical
value $\hat{u}_{2f}=1$ as long as the relative 
velocity ahead of the shock front, $u_D < \hat{u}_{1s} < u_R$. 

Thus, during the phase of the oscillation when the HII boundary is moving
inward and the upstream relative velocities are within this range, $u_D < \hat{u}_{1s} < u_R$, the density
of the ionized gas that enters the HII region decreases less than it does when the single ionization
front structure is allowed. After a crossing time,
this gas, now near the inner boundary, dominates the number of recombinations in the integral in
equations  \ref{frontv0} - \ref{frontv2} and
drives the outward motion of the ionization front less than it would 
without the transition to a double-front discontinuity. 
Thus if the initial conditions allow, the growth of the amplitudes of the oscillation is limited by the development of the
double-front discontinuity. 

The transition from an R-type to a D-type ionization front may not be dramatic. In our ODE
solution, we find that
the front may switch back and forth between critical and sub-critical several times per oscillation cycle
suggesting that a dense post-shock neutral layer does not necessarily build up and persist, at least not for 
the initial conditions of our example solution.

To continue the explanation, we compare the evolution of the ionization front within a Bondi accretion flow 
with two different models.
First, under the right conditions a nonlinearly damped harmonic oscillator can lead to conditionally stable 
oscillations.  
Second, the ionization front in the textbook model for the pressure-driven expansion of an HII
region within a uniform density gas also transitions from R-type to D-type, but
the conditions at the HII region boundary are controlled in a different way  \citep{Spitzer1978,Shu1991}.

\subsubsection{Comparison with a non-linearly damped harmonic oscillator}
While the evolution of the ionization front to conditionally stable oscillations 
is similar to that of a forced and damped harmonic oscillator leading to a limit cycle, 
the equation for the position and
velocity of the ionization front is different. The equation for a harmonic oscillator has the general form,
\begin{equation}\label{harmonic}
\ddot{x} + b(x)\dot{x} + cx = f(t) ,
\end{equation}
where $b,c$, and $f$ are arbitrary constants or functions.
If the damping term is suitably non-linear, in particular, stronger at the extrema of the oscillations, 
the evolution may lead to a classic limit cycle with conditionally stable oscillations
as in the well-known van der Pol oscillator.
In contrast, equation \ref{frontv1} for the velocity of the ionization front is of the form,
\begin{equation}\label{genfront1}
\dot{x} = g(x,t) 
\end{equation}
with a function $g$.
The dependence on position 
is seen in the upper limit of the integral in equation \ref{frontv1} and also in the velocity of the neutral flow at the position of the front.
The time dependencies result  from the advection of the density across the HII region and the  position of
the front.
The time derivative of equation \ref{frontv1} for the acceleration of the front is of the form,
\begin{equation}\label{genfront2}
\ddot{x} = h(\dot{x},t) .
\end{equation}
This is of course different from that of a harmonic oscillator (equation \ref{harmonic})
in that the function $h(\dot{x},t)$
depends only on time and the time derivative of $x$. In particular, the restoring force proportional
to $x$ is missing.  A simple ODE for the acceleration of the ionization front would be of the type, 
\begin{equation}
\ddot{x} + h(t)\dot{x} = 0,
\end{equation}
with a function $h$.
This type of equation can be solved with an integrating factor,
\begin{equation}
\mu(t) = \exp{\int h(t)dt}
\end{equation}
with the argument of the exponent setting a time-dependent  damping or amplification time scale.
Of course, owing to the complexity of the equations for accretion, it is not possible to write a closed-form equivalent 
of the function, $h$, and therefore
the differential equations for accretion cannot be solved in this way. Nonetheless, because equations \ref{harmonic} and \ref{genfront2}
are not of the same form, the dynamics of two-phase accretion dynamics are different from those of a nonlinearly damped harmonic
oscillator.

\subsubsection{Comparison with pressure-driven expansion}

In the textbook model for pressure-driven expansion of an HII region into a uniform, static gas \citep{Spitzer1978,Shu1991}, 
once the velocity of an R-type ionization front slows to its critical value, the ionization front
transitions to a D-type with a preceding isothermal shock. In this model, the velocity of
the front is consistent with ionization equilibrium as it is in the model for ionized accretion. However, the 
subsonic, pressure-driven expansion of the HII region determines the gas velocities across the shock and ionization fronts. In effect, the
jump conditions depend on the velocities from the inside out. 
In contrast, in ionized accretion, the gas within the HII region is
flowing inward, almost in free-fall, and hydrodynamic pressure is not effectively transmitted upstream. The jump conditions
depend on the velocities from the outside inward with the hydrodynamic pressure 
within the thin shell of post-shock neutral gas regulating the
speed of the isothermal shock. 
In both models, when $u_{1f} < u_R$ and a double-front structure is required, the result
is the same -- the gas enters the HII region at the ionized sound speed with respect to the velocity of the ionization front.

\section{Numerical Hydrodynamic Simulations}\label{NHS}

With the method of characteristics, we have described solutions of the coupled ODE's for the ionized gas that result in damped 
oscillations and conditionally stable oscillations for relative velocities above the R-critical limit, $\hat{u}_{1f} = u_R$, or
as an approximation, not much below the limit. Following the same method, we could extend the ODE model to 
include relative velocities arbitrarily below the R-critical limit by adding an additional zone to our model
to calculate the hydrodynamic evolution within the thin shell of post-shock neutral gas ahead of the ionization front when this shell develops as
a result of the transition from an R-type to a D-type ionization front. This would be a second IVP to be solved as a BVP
with the ionization front as the inner boundary and the shock as the outer boundary. 
However, with sufficient understanding of the dynamics gained through the
simplified physics described by the coupled ODEs of the two-zone model, a better course may be a change of method to
follow the evolution with a finite-volume numerical hydrodynamical simulation.

\subsection{Code}

The hydrodynamical evolution for the full accretion flow, neutral and ionized,
is followed using the open source code \emph{Pluto} version 4.1 \citep{Mignone2007} 
configured for one-dimensional (1D) simulations assuming spherical symmetry.
The ionization-radiation transport is solved using the ray-tracing module \emph{Sedna} \citep{Kuiper2020,Kuiper2018}. 
From the range of hydrodynamics solvers available in \emph{Pluto}, we use the hllc scheme (Harten - Lax - Van Leer scheme including the contact discontinuity).
With a WENO3 interpolation scheme in space and a Runge-Kutta 3 interpolation in time, the hydrodynamics are accurate to 3rd order in space and time.
The thermal equation of state of the gas is locally isothermal, with the actual gas temperature as a function only of the ionization fraction, $\epsilon$,
\begin{equation}
\label{eq:2T}
T_\mathrm{gas} = \epsilon \times T_\mathrm{ion} + (1-\epsilon) \times T_\mathrm{neu}
\end{equation}

\subsection{Initial Conditions}

Our initial model is chosen to be similar to that of  \citet{Vandenbroucke2019} with a fixed central mass, 
$M = 18 \mbox{ M}_\odot$, neutral and ionized gas temperatures, $T_1 = 500 \mbox{ K}$,  $T_2 = 8000 \mbox{ K}$
and the density at infinity normalized to $\rho_\infty = 10^{-19} \mbox{ g cm}^{-3}$.
The neutral and ionized sound speeds are $a_1=2.03$ and $a_2 = 11.5$ kms, respectively. 
The Bondi radii ($GM/(2a^2)$) for the two phases are $R_{B1}=1942$ and $R_{B2}=60.5$ au, respectively.

At the outer boundary of the computational domain, $R_\mathrm{max} = 100 \mbox{ au}$, 
we set the gas mass density and gas velocity to the analytical solution of the Bondi accretion flow, 
specifically $\rho_\mathrm{gas} \approx 2 \times 10^{-17} \mbox{ g cm}^{-3}$ and $v_\mathrm{gas} \approx -16.6 \mbox{ km s}^{-1}$.
At the inner boundary, $R_\mathrm{min}$,  we use zero-gradient boundary conditions, i.e.~mass, 
momentum, and energy are allowed to stream freely across the boundary. The spatial resolution of the grid is uniform with $\Delta r = 1/30 \mbox{ au}$. 
We ran simulations with two choices for $R_\mathrm{min} = 10$ (same as \citet{Vandenbroucke2019}), and also at $1$ au.

\begin{figure*}
\begin{tabular} {p{3.0in}c}
\includegraphics[width=3in]{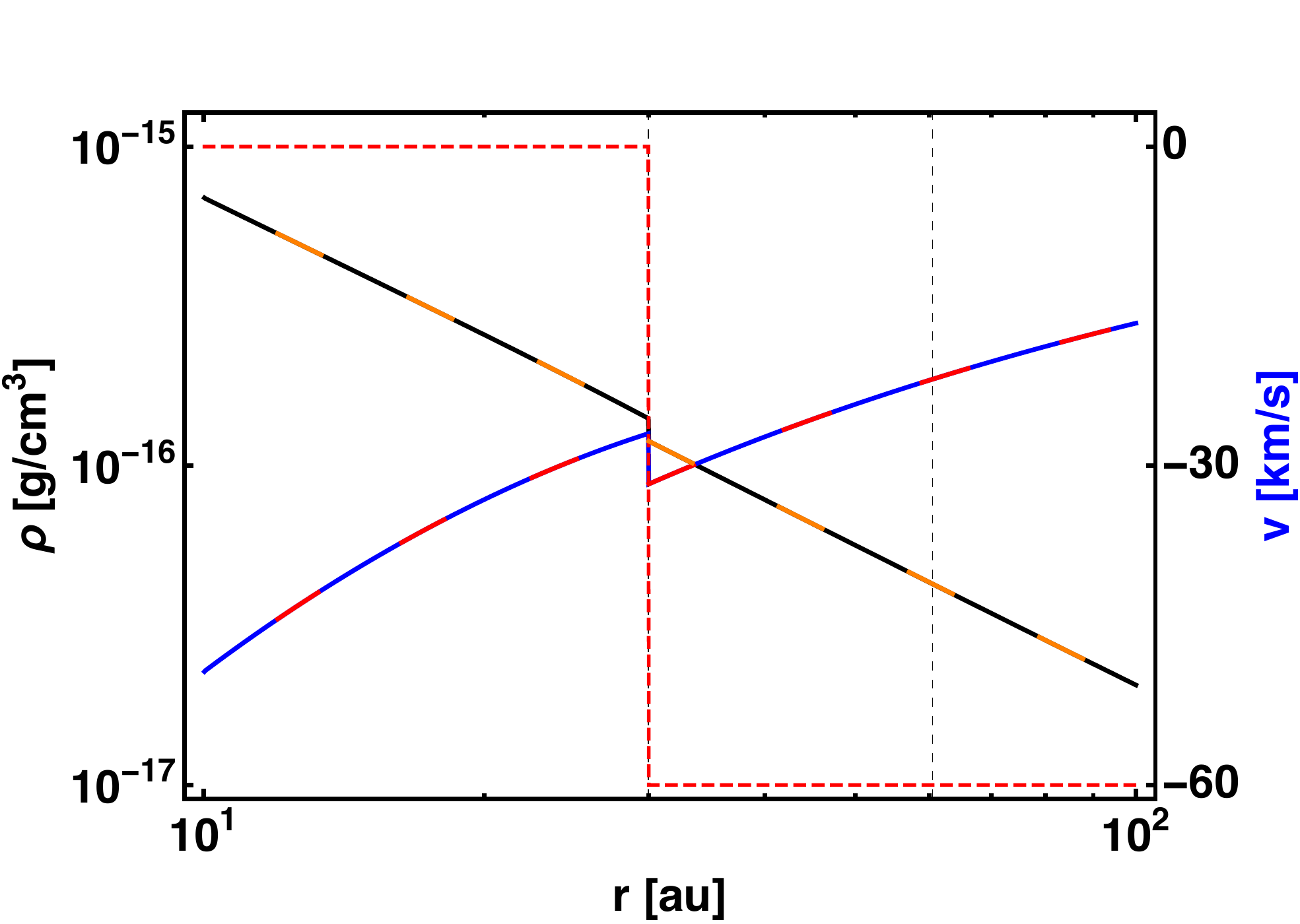}
&
\includegraphics[width=2.675in]{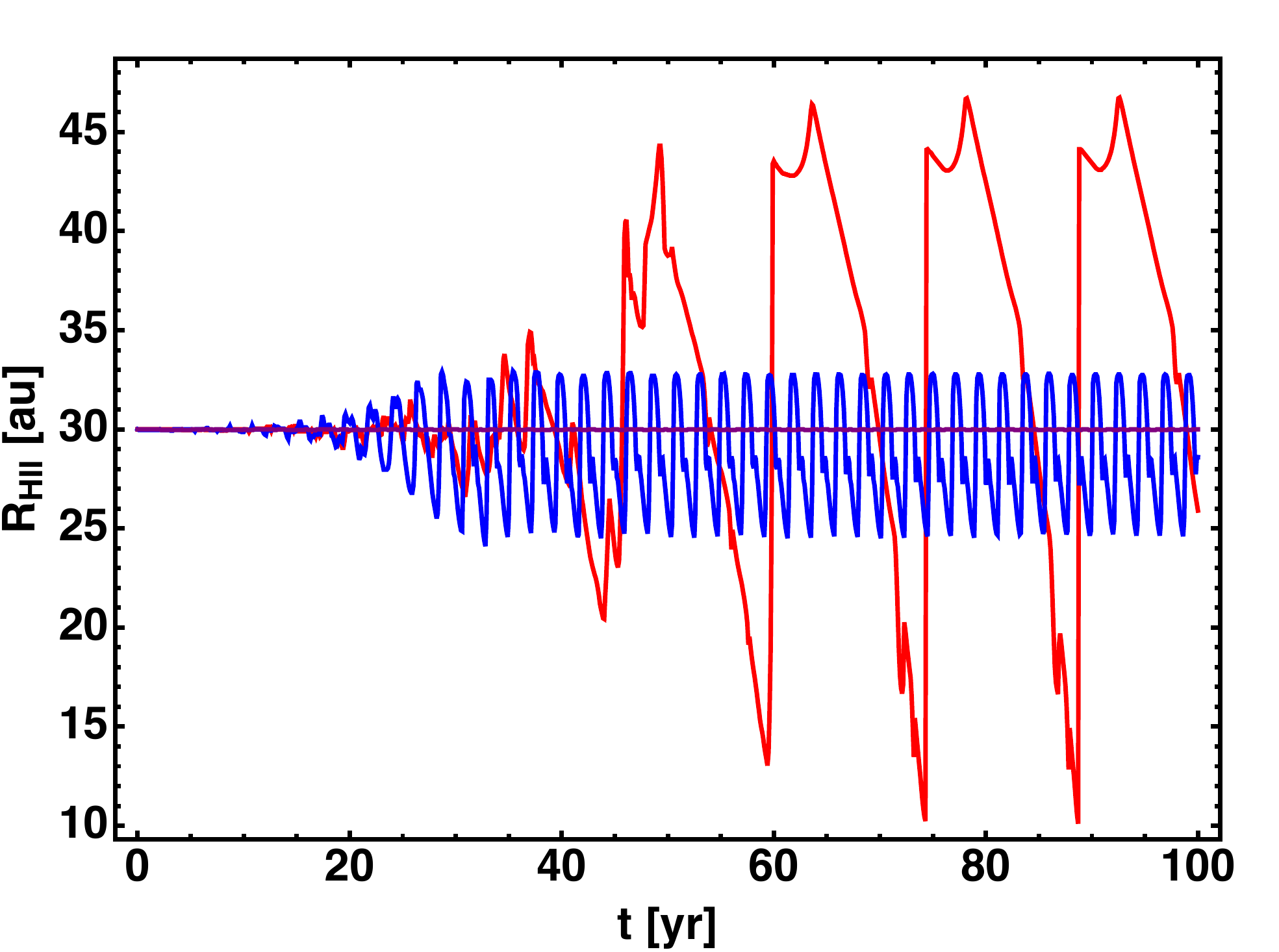}
\end{tabular}
\caption{
{\it Left}: Initial steady-state solution of the ionized Bondi accretion flow.
Gas mass density (left vertical axis)  as function of radius from the numerical simulation is shown as a black line overlain by an orange long-dashed line corresponding to the analytical solution.
Gas velocity (right vertical axis) as function of radius from the numerical simulation is shown as a blue line overlain by a red long-dashed line corresponding to the analytical solution.
The ionization fraction (dashed red line), jumps from 100\% to 0\% at the ionization front.
Vertical dashed lines denote the location of the initial ionization front at $30 \mbox{ au}$ and the Bondi radius with respect to the ionized gas temperature at $\approx 61.5 \mbox{ au}$.
Two movies showing the evolution of gas mass density and gas velocity from the uniform initial condition toward the steady-state solution of the ionized Bondi accretion flow are available online. A first movie shows the evolution from the uniform initial conditions to the steady-state solution of the neutral Bondi accretion flow. A second movie shows the further evolution after the addition of the ionization front. In the movies, dashed lines denote the analytic solutions for the neutral and ionized flow, respectively.
{\it Right}: Evolution of the location $R_f$ of the ionization front. The position of the ionization front as a function of time is shown from the initial conditions ({\it left} panel) for
the exact equilibrium (horizontal black line at 30 au) and
for a slight offset from equilibrium (red line) achieved by reducing the constant ionizing luminosity by $-0.06\%$ of
its equilibrium value. Also shown is the time evolution but with the ionized gas temperature  reduced by a factor of two to 4000 K (blue line).
Four movies showing the temporal evolution of the gas mass density and gas velocity profiles for ionized gas temperatures of 8000 K (equilibrium and nonequilibrium), 4000 K, and 2000 K are also available online.
}
\label{step2}
\end{figure*}

To develop an initial accretion flow in equilibrium,  we start with a uniform density, $\rho_{gas}$, equal to the value at 100 au given above and 
a uniform temperature, $T_1$, given above. 
As a first step, we let the system evolve without ionization under the gravitational influence of the constant central mass 
until the flow arrives at the single-phase Bondi transonic solution.
In a second step, we add an ionization front at a fixed position  (30 au and 12 au respectively in our two examples discussed below). 
The velocity of the neutral flow along with the gas temperatures
in the two phases define the jump conditions at $R_f$.  
We evolve this system hydrodynamically (without radiation transport) until another steady state is reached. 
The ionizing luminosity required for equilibrium is then derived from the density profile inside 
of the ionization front along with the neutral flux through the front. We now have the initial equilibrium state (figure \ref{step2}) for the 
fully time-dependent coupled hydrodynamic and photoionization-radiation transport problem. 

\subsection{Time-Dependent Evolution}
 
 We allow the position of the front to move
 in response to the ionizing flux at the front  which is calculated as the solution of the radiation transport problem. As shown in figure \ref{step2},
 the position of the front does not change from the initial equilibrium. However, the equilibrium is not stable.
 For example, if the ionizing luminosity is changed by $-0.06\%$ and then held constant, the position of the front develops
 oscillations, shown in our figure \ref{step2}, nearly identical to those found by \citet{Vandenbroucke2019}, their
 figures 6 and 8. In this example, the front periodically reaches the inner  boundary resulting
 momentarily in a fully neutral accretion flow. The HII region immediately 
 re-establishes itself with the position of the front in ionization equilibrium with the density profile of the
 neutral steady-state transonic Bondi solution. However, now the system is not in hydrodynamic equilibrium because 
 compression through the front requires an ionized density profile consistent with a different energy constant, $\mathscr{L}$ 
 (equation \ref{energyconstant}). 
 
 This behavior is not universal but dependent on the initial conditions. For example, reducing
 the ionized gas temperature to 4000 K results in a solution that is conditionally stable in that the maximum amplitude of
 the oscillations is limited, and the position of the front
 oscillates closely around the equilibrium position as seen in our figure \ref{step2}. If the ionized gas temperature is further reduced to
 2000 K, the solution remains essentially the same as the equilibrium solution. The lower ratio of ionized to neutral sound speeds
 weakens the compression across the ionization front which ultimately lengthens the response time of the ionization. In our ODE model,
 the relevant
 equations are \ref{frontv0} -- \ref{frontv2}.
 
Our non-dimensional ODE solution indicates that the differences in evolution
are due to the differences in the compression of the gas at the HII region boundary which is
set by the temperature ratio. (Our non-dimensional solutions shown in figures \ref{position1} and \ref{position_detail} have a temperature ratio of 100.)  For any
temperature ratio, solutions may be stable, conditionally stable, or unstable according to the initial position of the front with respect to the
temperature-dependent Bondi radius. For lower temperature ratios, stable solutions are found within a wider  range in the position of the front.

We next ran two series of simulations inserting a density perturbation in the neutral flow upstream of the ionization front.
Each series includes perturbations of different amplitudes, one per simulation, with multiplicative factors between 10 and 0.001 
 of the neutral steady-state transonic density.
 The density perturbations all have a width of 5 au. The initial (equilibrium) position of the ionization front is 12 au, $x_c/x_f = 1/5$. 
 The inner boundary is 10 au in the first series and 1 au in the second.
 The ionized and neutral temperatures are 8000 K and 500 K respectively, as in the original example.
  
\begin{figure*}
\begin{tabular} {p{3.0in}c}
\includegraphics[width=3.0in]{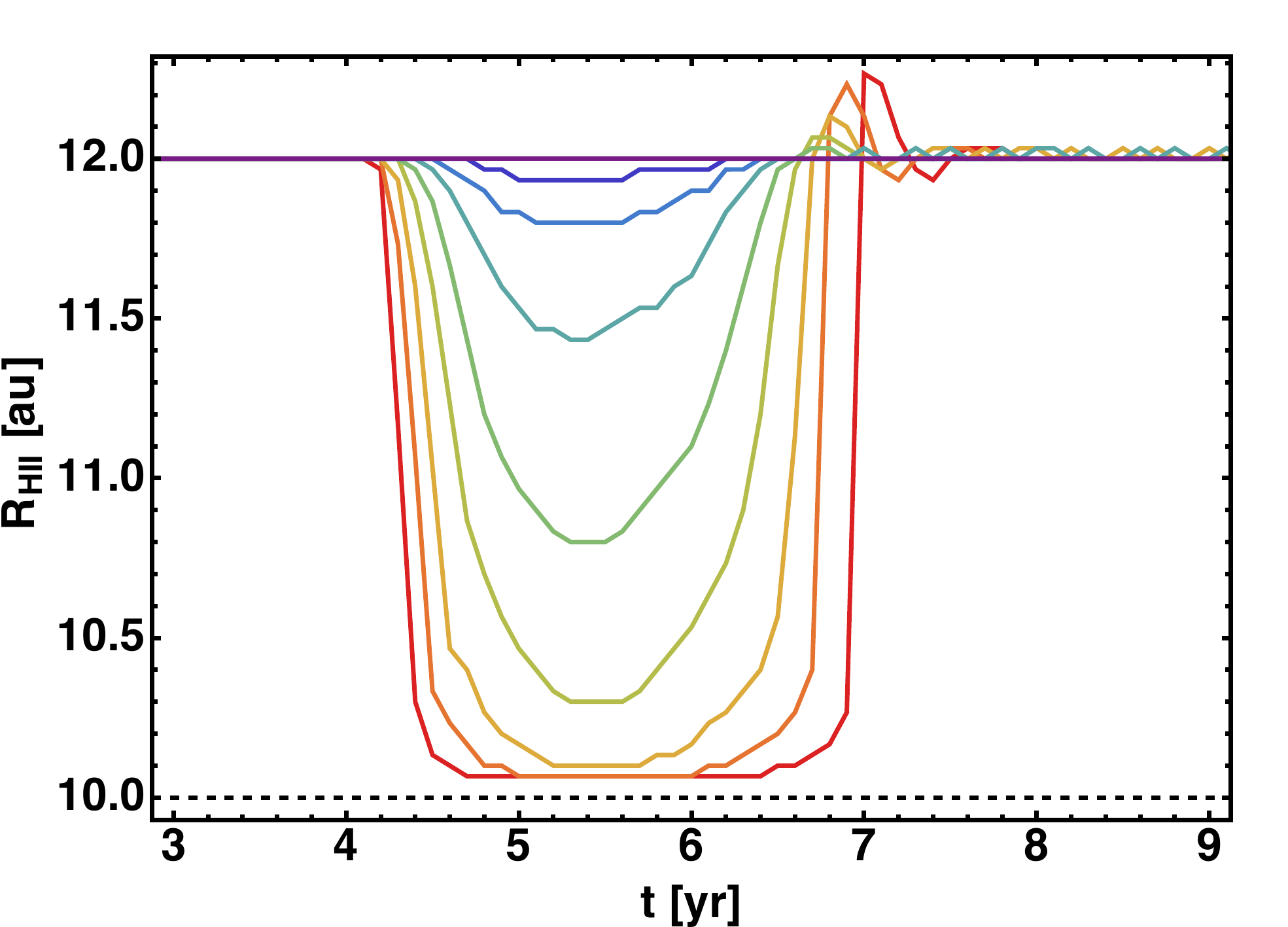}
&
\includegraphics[width=3in]{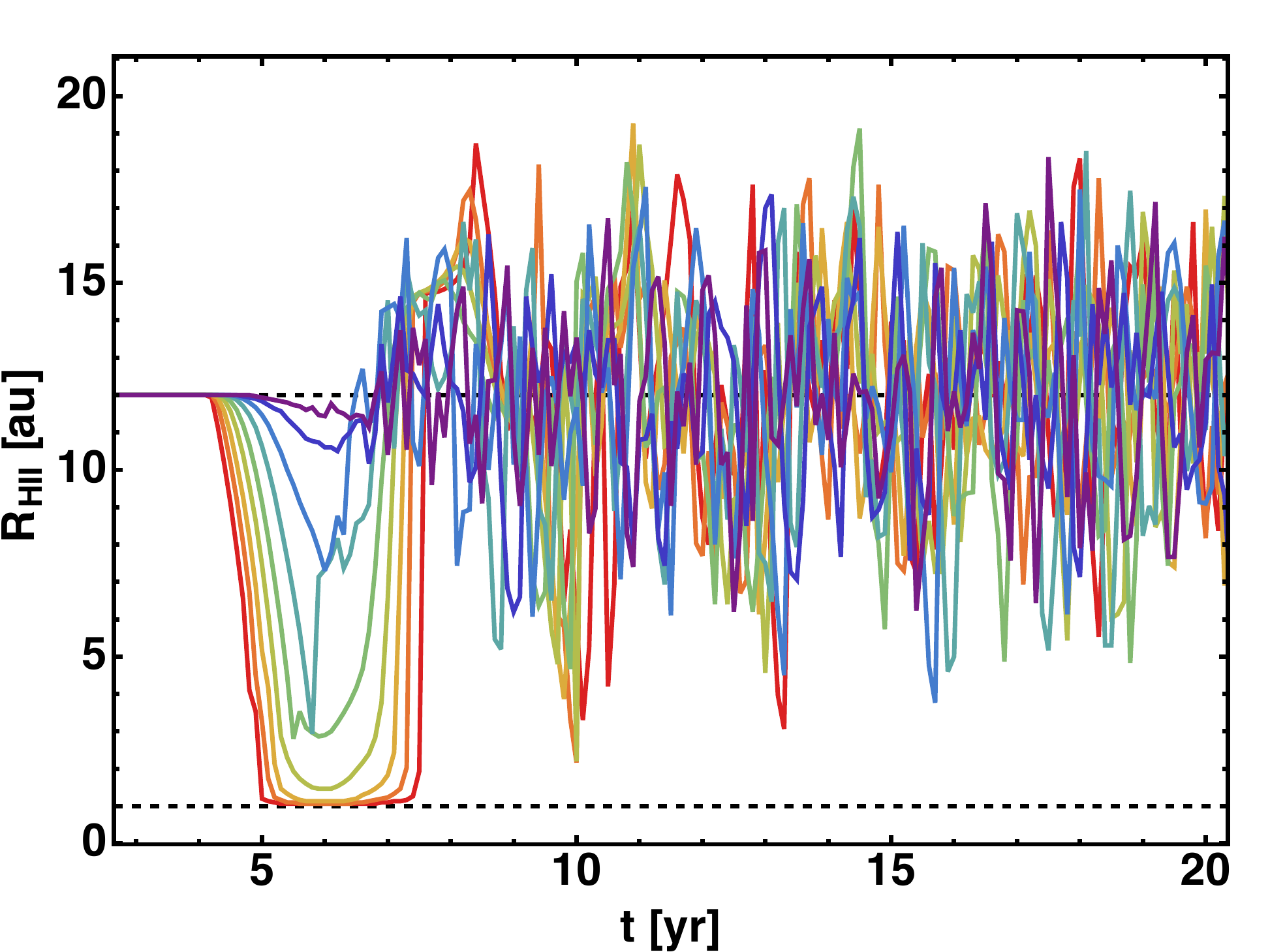}
\end{tabular}
\caption{
Temporal evolution of the location $R_f$ of the ionization front, initially located at $12 \mbox{ au}$ for nine different initial 
 perturbation strengths, 10.0, 3.0, 1.0, 0.3, 0.1,0.03,.0.01, 0.003, 0.001, times the steady-state transonic density, 
 shown as colors orange, yellow, light green, cyan blue, light blue, dark blue, and purple, respectively. The dashed
 horizontal line at 12 au shows the initial (equilibrium) position of the ionization front. The inner boundary is 10 au in the example shown {\it left},
 and 1 au in the example shown on the {\it right} indicated by dashed lines. 
 Two movies showing the temporal evolution of the gas mass density and gas velocity profiles are also available online (perturbations of 0.1 and inner boundaries of 10 and 1 au, respectively).
}
\label{perturbed}
\end{figure*}

 Figure \ref{perturbed} shows the results. Density perturbations of any amplitude result in oscillations of the position of the front. 
 Density perturbations of larger amplitude result in an initial collapse of the HII region through the inner boundary. The collapse occurs
 as the initial perturbation is advected through the ionization front and the density-dependent recombination rate limits 
 the number of ionizing photons reaching the front. Once a large amplitude
 perturbation has been mostly advected through the inner boundary, the HII region is re-established, and the position of the front then
 oscillates around its equilibrium value.
With the inner boundary at 
 10 au, the oscillations are damped resulting in a stable solution. With the inner boundary at 1 au, 
 the oscillations are conditionally stable with limited growth in amplitude.  Even after long evolutionary times (100 yr),
  these conditionally stable oscillations
 never result in subsequent collapse of the HII region. In these examples, the difference between the stable and conditionally stable behavior
 is due to
 the lower degree of spherical compression with a larger inner radius resulting in more stable behavior. If our non-dimensional
 ODE examples (figures \ref{position1} and \ref{position_detail}) 
 were scaled to correspond to the dimensional initial conditions of these numerical simulations, the inner boundary 
 in the ODE solutions would be 9.7 au.

\section{Astrophysical Implications}

The astrophysical applications of the instabilities in spherical ionized accretion may be limited because of the  dependence of the instability on
the spherical geometry of Bondi accretion flow and specifically on the spherical convergence of the flow. 
Astrophysical accretion flows are generally flattened in geometry due to the conservation of angular 
momentum. While the density in these two-dimensional, rotationally-flattened flows is also expected to increase at smaller radii, the 2D geometry 
lacks the spherical compression of the 1D model because the compression can be relieved by expansion of the disk scale height.
Oscillations seen in the spherical Bondi problem are not expected nor seen in 2D and 3D numerical simulations of
rotationally-flattened accretion flows with ionization \citep{Kuiper2018, Sartorio2019}.

\section{Conclusions}

The spherical convergence in Bondi accretion causes the density to increase
rapidly at smaller radii, approximately as $r^{-1.5}$ (free fall). In a two-phase accretion flow, because the recombination rate
increases as the square of the density, the effective opacity of the HII region and therefore the
ionizing flux at the front is controlled predominately by the higher density region at smaller radii.
The ionizing flux at the front controls the velocity of the front which in turn determines the
compression of the flow through the HII region boundary, and with a time lag due to advection, the density
at smaller radii. The offset of the phases of oscillations in the position and velocity of the front may lead to either damped
or growing oscillation amplitudes. Amplitude growth may be limited by the transition of the ionization front
from R-type to D-type and back as the transition limits the velocity of the front.

\section*{Acknowledgements}
RK acknowledges financial support via the Emmy Noether and Heisenberg Research Grants funded by the German Research Foundation (DFG) under grant no. KU 2849/3 and 2849/9.

\section*{Data Availability}
No new data were generated or analysed in support of this research.

\bibliographystyle{mnras}
\bibliography{final-2021-10-14} 





\bsp	
\label{lastpage}
\end{document}